\documentclass[12pt,preprint]{aastex}

\newcommand{\xsrc}{\mbox{XTE~J1908$+$094~}}

\begin{document} 

\title{Long Term Spectral and Timing Behavior of the Black Hole Candidate \xsrc} 

\author{
Ersin {G\"o\u{g}\"u\c{s}}\altaffilmark{1,2}, 
Mark~H.~Finger\altaffilmark{2,3}, 
Chryssa Kouveliotou\altaffilmark{3,4}, 
Peter~M.~Woods\altaffilmark{2,3},
Sandeep~K.~Patel\altaffilmark{3,5},
Michael~Rupen\altaffilmark{6},
Jean~H.~Swank\altaffilmark{7},
Craig~B.~Markwardt\altaffilmark{7},
Michiel~van~der~Klis\altaffilmark{8}
}

\altaffiltext{1}{Sabanc{\i} University, Orhanl{\i}-Tuzla 34956 {\.I}stanbul, Turkey}  
\altaffiltext{2}{Universities Space Research Association} 
\altaffiltext{3}{National Space Science and Technology Center, 320 Sparkman Dr.
Huntsville, AL 35805, USA}
\altaffiltext{4}{NASA Marshall Space Flight Center, SD-50, Huntsville, 
AL 35812, USA} 
\altaffiltext{5}{National Research Council Fellow}
\altaffiltext{6}{NRAO, Socorro, NM 87801, USA}
\altaffiltext{7}{NASA Goddard Space Flight Center, Greenbelt, MD 20771, USA}
\altaffiltext{8}{Astronomical Institute ``Anton Pannekoek'' and CHEAF,
University of Amsterdam, 403 Kruislaan, 1098 SJ Amsterdam, NL} 

\authoremail{Ersin.Gogus@msfc.nasa.gov} 

\begin{abstract} 

We present the long term X-ray light curves, detailed spectral and
timing analyses of XTE J1908+094 using the Rossi X-ray Timing Explorer 
Proportional Counter Array observations covering two outbursts in 2002
and early 2003. At the onset of the first outburst, the source was found
in a spectrally low/hard state lasting for $\sim$40 days, followed by 
a three day long transition to the high/soft state. The source flux 
(in 2$-$10 keV) reached $\sim$100 mCrab on 2002 April 6, then decayed rapidly.
In power spectra, we detect strong band-limited noise and varying low-frequency 
quasi periodic oscillations that evolved from $\sim$0.5 Hz to $\sim$5 Hz 
during the initial low/hard state of the source. 
We find that the second outburst closely 
resembled the spectral evolution of the first. The X-ray transient's 
overall outburst characteristics lead us to classify XTE J1908+094 as a 
black-hole candidate. Here we also derive precise X-ray position of the
source using Chandra observations which were performed during the 
decay phase of the first outburst and following the second outburst.

\end{abstract} 


\section{Introduction} 

The X-ray transient source \xsrc was serendipitously discovered on 2002 
February 19 during
scheduled Rossi X-ray Timing Explorer (RXTE) Proportional Counter Array (PCA) 
observations of a Soft Gamma Repeater, SGR 1900+14 (Woods et al. 2002). 
Subsequent RXTE/PCA scanning observations of the region allowed the localization
of the new source to RA: 19h08m50s, Dec: +09$^\circ$22$\arcmin$ 30$\arcsec$ 
with an accuracy of 2$\arcmin$. 
This placed the new source about 24$\arcmin$ away from SGR 1900+14. Based on 
Very Large Array observations on 2002 March 21 and 22, a transient source 
was suggested as the radio counterpart candidate to \xsrc (Rupen, Dhawan 
\& Mioduszewski 2002a). Observations in the optical band on April 8 and 9 
revealed no new sources near the radio position (Garnavich, 
Quinn \& Callanan 2002), however Chaty, Mignani \& Israel (2002) 
identified a near infrared counterpart to the new source and concluded that 
\xsrc is in a low-mass X-ray binary system with a 
main sequence companion of spectral type later than K. Based on our 
preliminary analysis
of the RXTE/PCA observations we concluded that \xsrc is a new stellar mass
black hole candidate (Woods et al. 2002).

There are currently 18 dynamically confirmed and 20 candidate stellar mass black
hole systems in our Galaxy (see McClintock \& Remillard 2003 for a recent 
review). Most of them are characterized by occasional transient outbursts 
(X-ray novae; Chen, 
Shrader \& Livio 1997) as a result of sudden increase in the mass accretion 
rate possibly triggered by instabilities in the accretion disk 
(Cannizzo 1993, Dubus et al. 2001).
During outbursts, these systems generally undergo 
various changes in their spectral characteristics, usually  in conjunction  
with changes in their timing behavior (see e.g., Homan et al. 2001). 
The most common BH spectral states are {\it the low/hard state:} the spectrum is 
represented by a hard power law and usually accompanied by timing 
variability, and {\it the thermal-dominant (high/soft) state:} a blackbody 
appears in the spectrum as the power law component gets steeper and timing 
features get weaker or completely disappear (Tanaka \& Lewin 1995). 
Other spectral states characterized by more complicated 
spectral and timing properties are also observed (e.g., Homan et al. 2001).

In this study, we present the results of our spectral and timing analysis of 
the RXTE pointed observations of \xsrc covering two outburst episodes. Additionally,
we report observations with the Chandra X-Ray Observatory taken under 
Director's Discretionary Time (DDT). We describe our observations in \S 2, 
and we present detailed data analyses in \S 3. In \S 4 we discuss and 
compare the different states of the source to other black hole candidates.

\section{Observations} 

\begin{figure}[!b]
\centerline{\includegraphics[scale=.50]{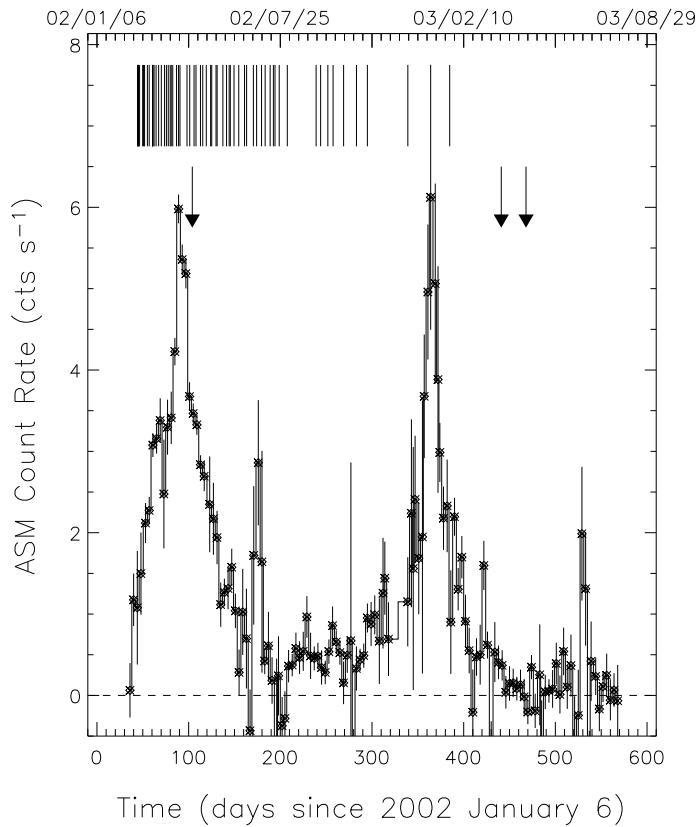}}
\vspace{0.2in}
\caption{\baselineskip =0.5\baselineskip 
The light curve of \xsrc as seen in 1.5 $-$ 12 keV with the RXTE/ASM. 
The vertical lines indicate the times of pointed RXTE observations of the 
source. The arrows denote the times of Chandra pointings. Observation times 
are relative to 2002 January 6 (MJD 52280). The calendar dates indicated above
are in YY/MM/DD format.
\label{fig:rates_obss}}
\end{figure}

In Figure \ref{fig:rates_obss}, we show the light curve of \xsrc in 
the 1.5$-$12 keV band observed with the All Sky Monitor (ASM) onboard RXTE. 
Pointed observations with RXTE, indicated by vertical lines, were made 
throughout two outburst episodes until 2003 January 31, although pointings 
were more frequent during the first outburst. In this study, we used the data 
from 58 pointed RXTE
observations (Obs \# 7 through 64 in Table \ref{tbl:1}) with a total exposure
time of 110.4 ks. In addition, we have 6 observations of SGR 1900+14 
from our monitoring program (Obs \# 1 $-$ 6 in Table \ref{tbl:1}), covering
the same portion of the sky $\sim$40 days prior to the first transient outburst of
\xsrc. We use these pointings to determine the background behavior from other 
sources in the field and the galactic ridge contribution as we will describe in \S 3.3. 
For convenience, we reference all times to 2002 January 6
(MJD 52280) in this paper.

Besides the ASM, there are two more instruments on RXTE: the PCA,  
an array of 5 nearly identical Proportional Counter Units (PCUs) which are 
sensitive to photon energies between $2-60$ keV, and the High Energy X-ray 
Timing Experiment (HEXTE) sensitive to photon energies between $15-250$ keV.
Here, we present the results of our PCA data analysis only. The HEXTE spectral 
data analysis resulted in large statistical uncertainties after the background 
subtraction, even during the phase when the hard spectral component 
dominates the source spectrum, and it is not presented.
 
A short (1.1 ks) Chandra Advanced CCD Imaging Spectrometer (ACIS) observation 
was performed on 2002 April 15. It took place during the relatively bright phase
of the outburst (day 99 in Figure \ref{fig:rates_obss}), resulting in heavy pile up. 
Nevertheless, we could still use the data to determine an accurate X-ray position of \xsrc. There are,
additionally, two publicly available  Chandra ACIS pointings performed on
2003 March 23, and 2003 April 19 for $\sim$ 4.7~ks each. We used these
observations to confirm our best location solution determined from the earlier
short pointing.

\section{Data Analysis and Results}

\subsection{Source Location from the Chandra/ACIS Data} 

Our 2002 Chandra observation was taken in the ACIS full frame timed 
exposure mode while the two subsequent observations were collected in the 1/8
subarray mode.
Initial standard processing of the data was performed by the Chandra 
X-ray Center (CXC).  We modified the standard processed data by using 
{\sl acis\_process\_events} (CIAO v2.3) 
to remove pixel randomization and to retrieve x-ray events 
flagged bad while attempting to remove cosmic ray events.  The data 
were filtered to exclude events with ASCA grades 1, 5, and 7, hot 
pixels, bad columns, and events on CCD node boundaries.  

To determine the location of \xsrc, we followed the method of Hulleman 
et al. (2001) and fitted the piled up image with an appropiate function.  
We also determined the  position from the two observations in 2003 
and found that they are consistent.  Since these data were not significantly 
piled up, we fitted a 2-dimensional Gaussian to the source location and
determined the source centroid.  We averaged all these locations and derived 
a (J2000) position of $\alpha=19^{\rm h} 08^{\rm m} 53^{\rm s}.07$, 
$\delta=+09^\circ 23\arcmin 04\arcsec.9$ for \xsrc\ (90\% error 
radius\footnote{http://asc.harvard.edu/cal/ASPECT/celmon} of $0.6\arcsec$).

\subsection{Light Curves, Hardness Ratios, Color-Intensity Diagram} 

\begin{figure}[!b]
\centerline{\includegraphics[scale=.50]{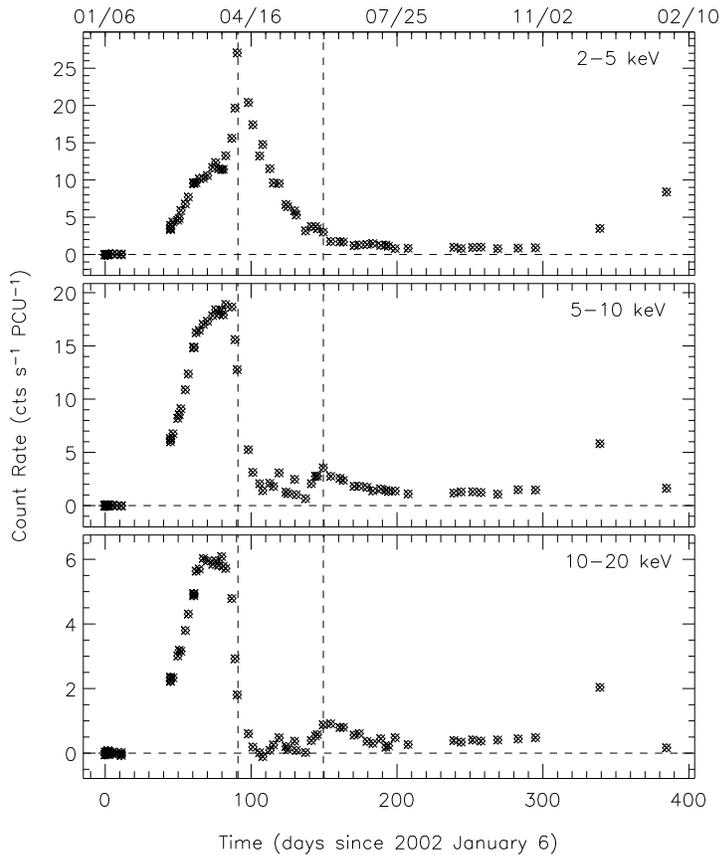}}
\vspace{0.2in}
\caption{\baselineskip =0.5\baselineskip
The light curves of \xsrc as seen in the 2$-$5, 5$-$10 and 10$-$20 keV 
ranges. Each point represents the averaged and background subtracted count 
rates of each RXTE/PCA orbit in a given energy interval. 
The dashed vertical lines 
indicate the times of spectral state transitions, from hard to soft ({\it left}),
and soft to hard ({\it right}). The calendar dates indicated above
are in MM/DD format.}
\label{fig:rates_all}
\end{figure}

In Figure \ref{fig:rates_all}, we show the long-term PCA light curves of 
\xsrc in the 2$-$5, 5$-$10 and 10$-$20 keV energy bands. 
During the first $\sim$16 days of the outburst 
(from day 44 to 60 in Figure \ref{fig:rates_all}), the rates in 
all 3 bands increased by a factor of $\sim$3. In the following $\sim$27 days,
the rise was gradual in the 2$-$5 and 5$-$10 keV bands, while the 10$-$20 keV 
rate remained constant.

Starting at day $\sim$87, the rate in the 2$-$5 keV range increased rapidly 
(by a factor of 2 over the next 3 days), and then decayed exponentially 
(e-folding time of 26.8 $\pm$ 1.2 days). During the time of the low energy peak
(indicated by the left dashed vertical lines in Figure \ref{fig:rates_all}),
the rates in both higher energy bands sharply declined. 

\begin{figure}[!t]
\vspace{-0.2in}
\centerline{\includegraphics[scale=.50]{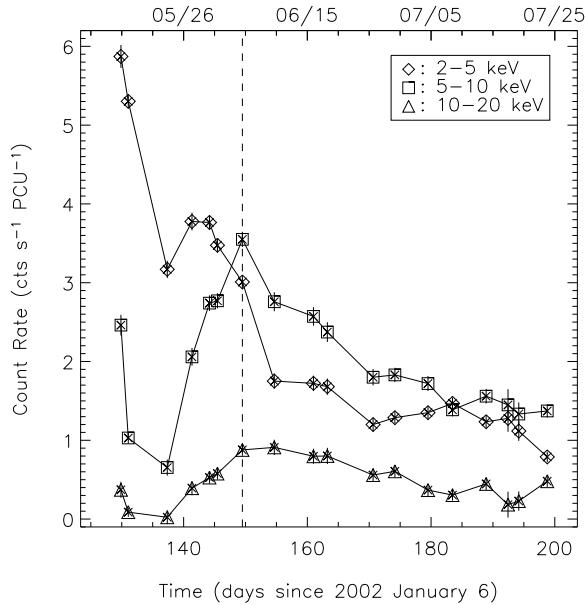}}
\vspace{0.2in}
\caption{\baselineskip =0.5\baselineskip
A zoom up of the \xsrc light curves around the time of the 
secondary peak in three energy bands. The calendar dates indicated above
are in MM/DD format.
\label{fig:rates_second}}
\end{figure}

About 55 days after the low energy peak, we see a secondary peak in 
all 3 energy bands, possibly a secondary maximum often seen in X-ray novae 
(Chen, Shrader \& Livio 1997).
A closer look at the secondary peak (shown in Figure \ref{fig:rates_second})
indicates that it evolves from soft to hard, with the 2$-$5 keV 
band peaking about 3 days earlier than the 5$-$10 keV energy band. It 
is noteworthy that soon after this secondary peak, the source 
transitioned into a harder spectral state (as marked in Figure \ref{fig:rates_all} 
with the right dashed vertical line; see also Figure \ref{fig:rates_second}).
 
Between days 170 and 295, the source intensity remained very low (the flux 
varied between 1.6 and 3.6 $\times$ 10$^{-10}$ erg s$^{-1}$
cm$^{-2}$), but significantly above the background value.
(see Figure \ref{fig:rates_obss} and \ref{fig:rates_all}). 
The last two PCA observations of \xsrc were during the second outburst 
episode which took place between days $\sim$330 and 410 (Figure 
\ref{fig:rates_obss}). The peak of the outburst as seen with the ASM is 
between these two observations. 

\begin{figure}[!t]
\vspace{-0.1in}
\centerline{\includegraphics[scale=.45]{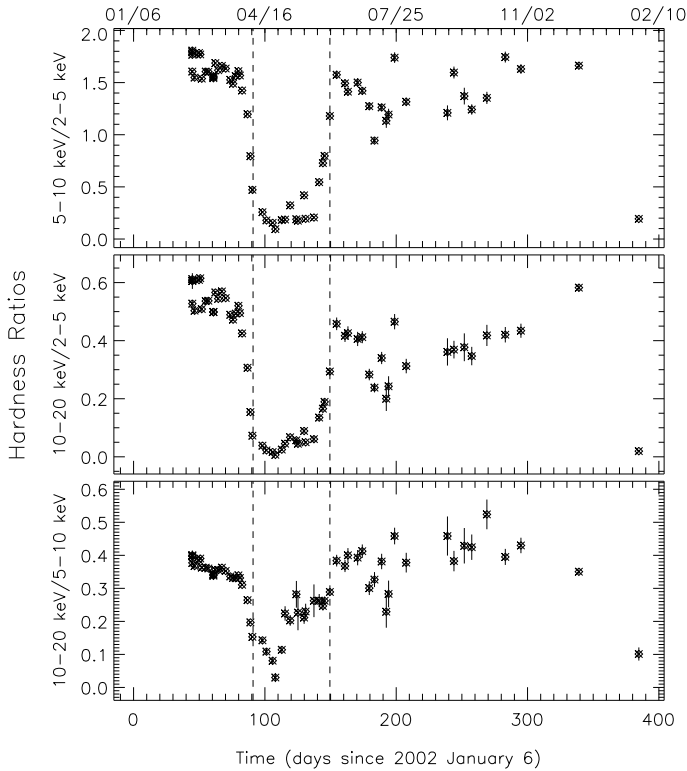}}
\vspace{0.2in}
\caption{\baselineskip =0.5\baselineskip
The evolution of the hardness ratios in \xsrc defined as the 
ordinate names of each panel. The calendar dates indicated above
are in MM/DD format.
\label{fig:hard_ratio}}
\end{figure}

Changes in the source spectral properties are also reflected in the hardness 
ratios as illustrated in Figure \ref{fig:hard_ratio}.
Here, the hardness ratios are defined as the ratio of the averaged (over each 
observation) count rates in the higher energy band to those 
in the lower energy band. Note that the evolution of the soft color
(Figure \ref{fig:hard_ratio}, top panel) and the hard color (Figure 
\ref{fig:hard_ratio}, middle panel) are quite similar. 
The rise in hardness after the dip coincides with the secondary peak shown 
earlier (denoted by the right dashed vertical line in Figure 
\ref{fig:hard_ratio}).

\begin{figure}[!t]
\vspace{-0.2in}
\centerline{\includegraphics[scale=.44]{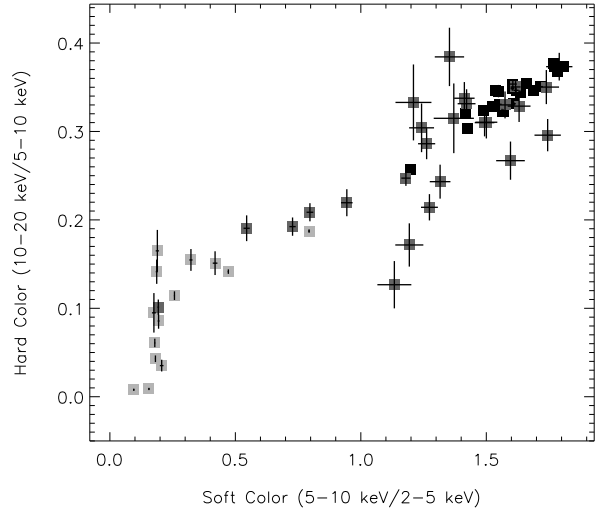}}
\vspace{0.1in}
\caption{\baselineskip =0.5\baselineskip
Color-color diagram of \xsrc. The hard color is the ratio of 
(background subtracted) rates in the 10$-$20 keV band to the 5$-$10 keV band 
for each observing window, and the soft color is the ratio of those in 5$-$10 
keV to the 2$-$5 keV band.\label{fig:hard_hard}}
\end{figure}

We present a color-color diagram in Figure \ref{fig:hard_hard}. 
The different colors of the symbols correspond to the spectral states of the source; 
black represents the early hard state ( days 44$-$90), and these points are,
therefore, concentrated at the upper right corner of the diagram. Light 
gray represents the soft state ( days 90$-$140) and dark gray 
are points from the subsequent low state (days 140$-$ afterwards). 

\begin{figure}[!b]
\centerline{\includegraphics[scale=.44]{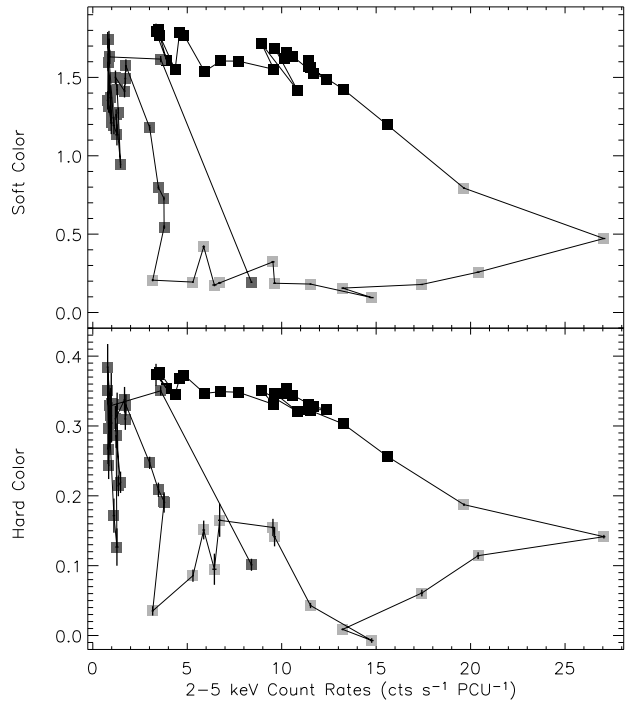}}
\vspace{0.1in}
\caption{\baselineskip =0.5\baselineskip
Color-Intensity diagram for the soft (upper panel) and the hard colors
(lower panel). \label{fig:hard_int}}
\end{figure}

In Figure \ref{fig:hard_int}, we show the color-intensity diagram for
both the soft and hard colors. Colors are identical to those in
Figure \ref{fig:hard_hard}. After the onset of the first outburst, both colors 
remained at high levels, then decreased quickly as the source underwent the
first state transition and looped in the clockwise direction. Notice 
that the color of first of the last two observations appears to be at the 
level of the initial hard state, and the color of the last observation is 
among those in the soft state.
Therefore, the second outburst, like the first, transitioned from a hard 
to a soft spectral state.

\subsection{Spectral Analysis} 

For each RXTE observing orbit, we extracted the PCA spectrum using Standard2 
data (129 channels accumulated every 16 s) collected from the top layer of all 
operating PCUs (to achieve the highest signal-to-noise ratio), excluding PCU0 
(because of its higher background level caused by the loss of its propane 
layer). In selecting data we required the Earth 
elevation angle to be greater than 10$^\circ$ and the time to the nearest 
South Atlantic Anomaly passage to be greater than 30 minutes. A background 
spectrum was generated using the faint source background models 
provided by the PCA instrument team and {\it pcabackest} which is an FTOOLS 
utility. Spectral modelling was performed using XSPEC 11.2.0.

Due to the fact that the PCA is not an imaging instrument and it has a 
relatively large field-of-view ($\approx 1^\circ$ FWHM), the background 
subtracted PCA spectra of \xsrc still have some contamination from other 
X-ray sources in the field and from the galactic ridge. To determine the 
spectral shape and the intensity 
level of this contamination, we generated background subtracted spectra from the
PCA observations performed $\sim$ 40 days before the outburst (Obs \# 1 $-$ 6 in
Table \ref{tbl:1}) using the same methodology described earlier. We found that
all these spectra were well fitted by a power law ($\Gamma$ = 1.8) plus a 
Gaussian line (E$_{\rm cent}$ = 6.68 keV and E$_{\rm width}$ = 0.45 keV) both attenuated
by the interstellar absorption (fixed at N$_{\rm H}$ = 1.8 $\times$ 10$^{22}$ 
cm$^{-2}$, as given by Valinia \& Marshall 1998). The model normalizations of 
all six spectral fits are consistent within errors, therefore we calculated
their weighted mean (by statistical errors).
We then included this absorbed power law plus Gaussian line model 
(all parameters are fixed at determined values) as a fixed term in the 
subsequent spectral fitting of the \xsrc.

The energy spectra of the black hole X-ray binaries are generally successfully 
fitted  by a two component model containing a multicolor disk blackbody 
(Mitsuda et al. 1984) plus a power law (Tanaka \& Lewin 1995). The former 
(or the soft) component is expected to originate mainly from the inner 
portions of the accretion disk (T$_{\rm disk} \propto$ R$^{-3/4}$), 
while the latter (the hard) component is believed to be 
due to Compton up-scattering of low energy seed photons from the disk by the 
energetic electrons (possibly) in  the hot corona around the inner disk.

\begin{figure}[!t]
\vspace{-0.2in}
\centerline{\includegraphics[scale=.50]{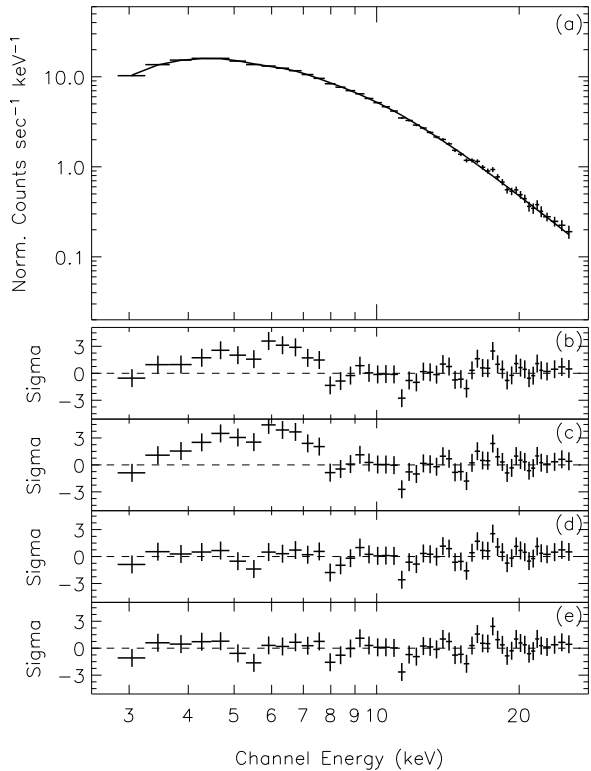}}
\vspace{0.1in}
\caption{\baselineskip =0.5\baselineskip
Representative spectral fit of a PCA observation during 
the early hard state of \xsrc (on day 62); (a) count spectrum and the 
best fit multicolor 
disk blackbody (MCD) + broad line + power law model all attenuated by the
interstellar absorption (solid line) (b) residuals of
the best fit single power law model ($\chi_{\nu}^2$=1.87), (c) residuals of 
the best fit MCD + power law model ($\chi_{\nu}^2$=2.66), (d) residuals of 
the best fit broad line + power law model ($\chi_{\nu}^2$=0.88), and 
(e) residuals of the best fit MCD + broad line + power law model
($\chi_{\nu}^2$=0.87).\label{fig:xte_spec0801}} 
\end{figure}

\begin{figure}[!t]
\vspace{-0.2in}
\centerline{\includegraphics[scale=.50]{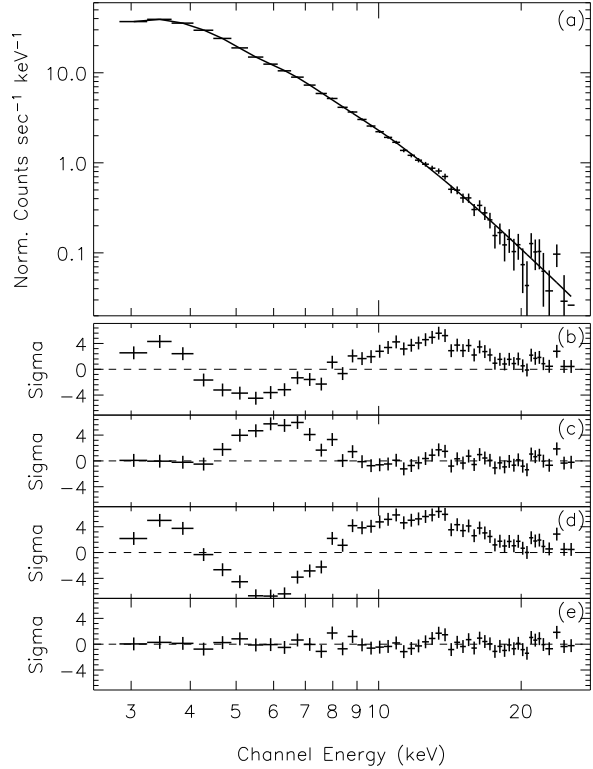}}
\vspace{0.1in}
\caption{\baselineskip =0.5\baselineskip
Representative spectral fit of a PCA observation during 
the intermediate state of \xsrc (on day 90); (a) count spectrum and the best fit multicolor 
disk blackbody (MCD) + broad line + power law model all attenuated by the
interstellar absorption (solid line) (b) residuals of
the best fit single power law model ($\chi_{\nu}^2$=8.26), (c) residuals of 
the best fit MCD + power law model ($\chi_{\nu}^2$=4.17), (d) residuals of 
the best fit broad line + power law model ($\chi_{\nu}^2$=14.50), and 
(e) residuals of the best fit MCD + broad line + power law model
($\chi_{\nu}^2$=0.70).\label{fig:xte_spec_1200}}
\end{figure}

Our initial fits to each \xsrc spectrum (over the energy range of 2.5$-$25 keV) 
with absorbed (N$_{\rm H}$ fixed at 2.5$\times$ 10$^{22}$ cm$^{-2}$ as reported 
by in't Zand et al. 2002) disk blackbody plus power law model during the early 
hard state of the outburst (from days 44 through 90) revealed a 
significant excess in the resulting residuals between 4.5 and 7 keV. 
Inclusion of a Gaussian line component to the model resulted in a broad line 
feature (see the two representative spectra in Figure
\ref{fig:xte_spec0801} and \ref{fig:xte_spec_1200} for the necessity of adding  
this component) and significant improvement in the goodness of fit to the 
acceptable level. Also during this state, the disk blackbody component was not
statistically required, nonetheless we kept this component to determine the
upper limits of the disk blackbody flux. Table \ref{tbl:2} lists the resulting
values for the spectral parameters. In all spectral fits, 
we obtained statistically acceptable $\chi_{\nu}^2$ values ranging 
between 0.51 and 1.19.

\begin{figure}[!t]
\centerline{\includegraphics[scale=.50]{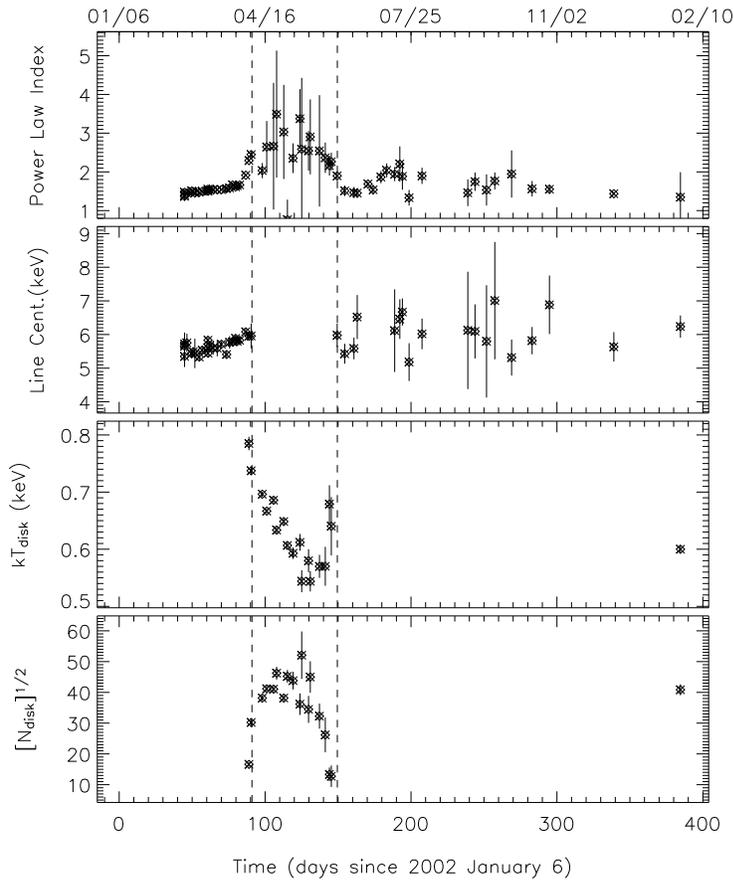}}
\vspace{0.2in}
\caption{Time history of the spectral model parameters. \label{fig:xte_spec}}
\end{figure}

We show the variations of the spectral model parameters in Figure 
\ref{fig:xte_spec}. During the low/hard state, the power law index remained 
fairly constant ($\Gamma$ $\sim$~1.5). The centroid of the broad line feature
appears to vary between 5.18 and 6.88 keV, with a statistically weighted 
average of 5.73 $\pm$ 0.09 keV. The width of the line ranges between 0.96
and 1.27 keV, with a weighted average value of 1.11 $\pm$ 0.31 keV. The
corresponding line equivalent widths range from 153 eV to 734 eV, with the 
majority being between 170 and 360 eV.

As the source transited into the soft state (first dashed line in Figure 9), the disk blackbody component 
appeared and became prominent, while the broad line feature disappeared.
The blackbody temperature clearly evolves during the soft state. The model
normalization, N$_{\rm disk}$ of the disk blackbody component is parametrized 
in terms of the inner disk radius, the distance to the source and 
the inclination angle, as N$_{\rm disk}$ = (R$^2_{\rm in,km}$/d$^2_{\rm 10kpc}$)
cos${\rm \theta}$. If the last two are assumed to not vary over the course of
the soft spectral state, there is some evidence for an early increase of the 
inner disk radius that remains relatively constant around $\sim$40 until 
day 130, and then decreases to its initial value. Meanwhile, 
the power law trend became steeper with an average value of $\Gamma \sim $2.5.

\begin{figure}[!t]
\centerline{\includegraphics[scale=.50]{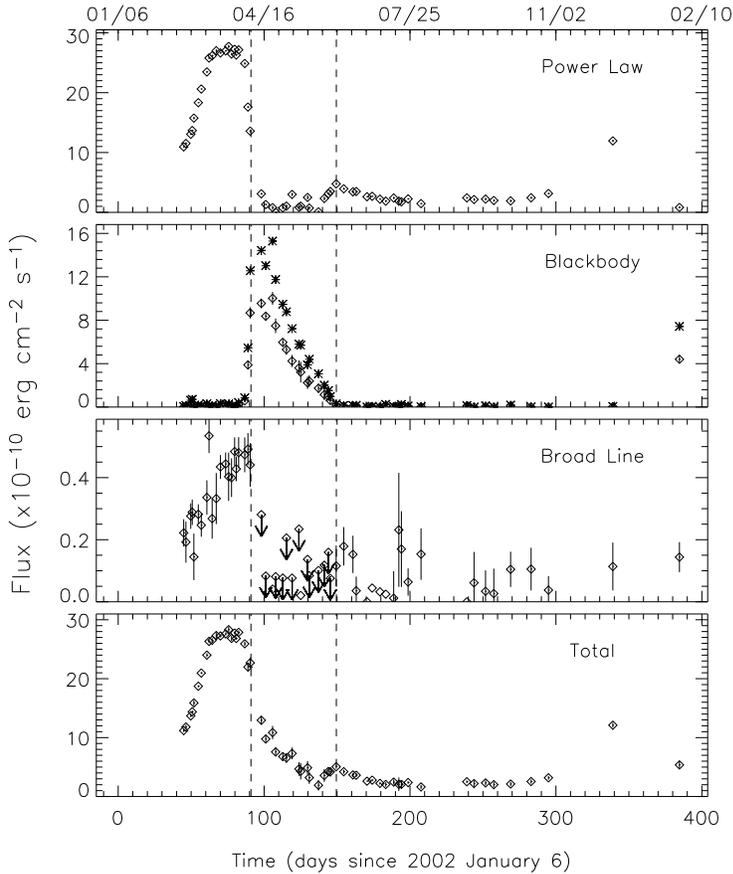}}
\vspace{0.2in}
\caption{\baselineskip =0.5\baselineskip
Time history of the absorbed fluxes of each model component 
integrated over 
the 2.5$-$25 keV
 range. The blackbody flux values integrated over the 
1$-$25 keV are shown with asterisks in the second panel from top. The
upper flux limits of the broad line feature indicate the 2$\sigma$ level.
\label{fig:xte_flux}}
\end{figure}

On day $\sim$ 149, soon after the secondary peak (second dashed line in 
Figure 9), the spectral transition into the low/hard state 
took place. Similar to the earlier hard state, the source spectrum 
is dominated 
by the power law component and the broad line feature again becomes observable,
although with relatively lower line flux, therefore, with larger uncertainties 
in the line model parameters (see also Figure \ref{fig:xte_flux}). 
 
The last two RXTE observations studied here (on days 339 and 384) 
captured a glimpse of the second outburst; the first of these pointings 
was during the rise of the X-ray intensity while the second was during its decline 
(see Figure \ref{fig:rates_obss}). Detailed spectral
analysis of these two observations show that the spectrum of the first one was 
hard, resembling the low/hard state; in contrast, the spectrum of the second was 
strongly dominated by the soft component, indicating the reappearance of
disk emission.

Figure \ref{fig:xte_flux} illustrates the time history of the fluxes of each
model component. Similar to the 10$-$20 keV light curve of the source (Figure 
\ref{fig:rates_all}, bottom panel), the power law flux increased over the
first 16 days into the outburst, then remained constant, forming the `knee' at
around day 60. There is an apparent correlation between the line and power 
law fluxes (especially during the first rising portion and around the secondary 
peak), which suggests that these two may be related.

In the energy range that the spectral fitting was performed (2.5$-$25 keV), 
we have been detecting only the higher energy tail of the blackbody spectrum
(kT $\sim$ 0.6$-$0.7 keV) whose significant portion lies between 1 and 2.5 keV.
To account for this, we have additionally integrated the blackbody
flux values over 1$-$25 keV (indicated by asterisks in Figure
\ref{fig:xte_flux}, second panel from top). We find that the absorbed 1$-$25
keV disk flux is $\sim$ 50\% larger than the flux in the 2.5$-$25 keV range.

\subsection{Timing Analysis} 

We computed power spectra  for each pointed observation using the PCA event
mode data (E\_125us\_64M\_1s). In the few cases where more than one PCU 
combination occurred during an observation, we made separate power spectra 
for each combination. More specifically, we generated light-curves 
from the event mode data using 31.25 ms binning, and divided them into 512 s 
segments, which were then Fourier transformed. The Fourier
powers were then averaged over the segments.

\begin{figure}[!t]
\vspace{-0.2in}
\centerline{\includegraphics[scale=.50]{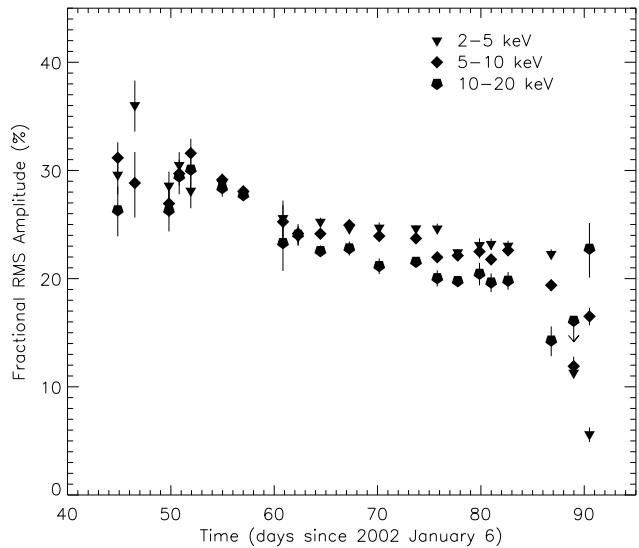}}
\vspace{0.0in}
\caption{\baselineskip =0.5\baselineskip
The fractional RMS amplitudes were determined for 3 energy ranges in 
frequencies between 0.002 and 5 Hz. The abrupt decline of the RMS amplitudes 
in all 3 energy ranges coincides with the spectral transition from the low/hard 
state to the high/soft state. \label{fig:xte_rmshist}} 
\end{figure}

To examine the overall evolution of the source noise during the
first outburst, we generated power spectra in three energy bands,
2$-$5 keV, 5$-$10 keV, and 10$-$20 keV. Figure \ref{fig:xte_rmshist} shows the 
fractional RMS amplitude in the 0.002-5 Hz frequency range for each energy band,
during the first outburst. The RMS amplitudes were normalized to the source 
count rates obtained by subtracting from the observed count rate the 
faint source model rate, and the predicted rate from the galactic 
ridge emission. There is no significant difference between
energy bands. The fractional variability drops slowly until day 87, 
after which it falls rapidly in coincidence with the
fall of the 10$-$20 keV flux, and the rapid
rise in the 2$-$5 keV flux. 
After the peak of the 2$-$5 keV flux on day 90, 
the source noise power is no longer detected.

\begin{figure}[!t]
\vspace{-0.2in}
\centerline{\includegraphics[scale=.50]{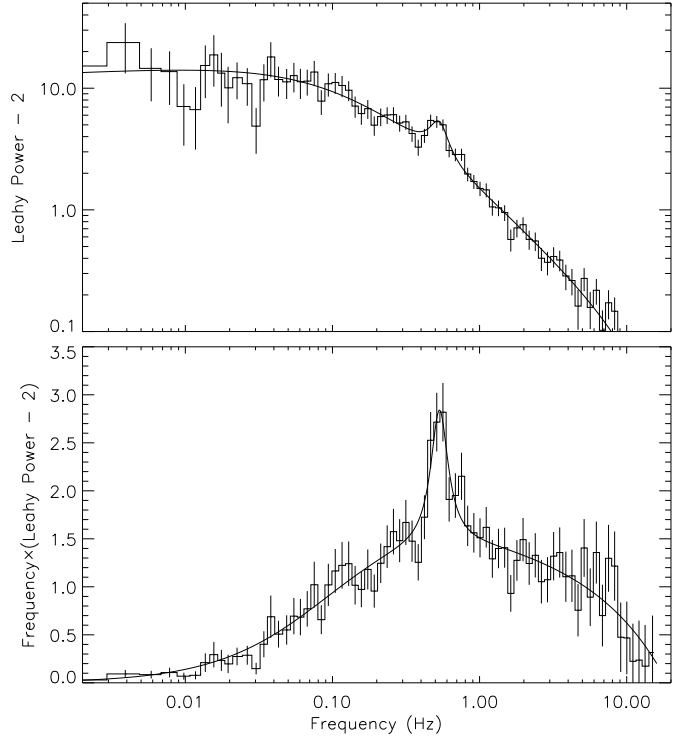}}
\caption{\baselineskip =0.5\baselineskip
(upper panel) Representative power spectrum of RXTE observations 
on day 54.9 (2002 March 1.9). (lower panel) Same as upper panel
multiplied by the frequency. Notice the QPO peak at $\sim0.5$ Hz.
\label{fig:xte_powspec}}
\end{figure}

Figure \ref{fig:xte_powspec} shows a representative power spectrum 
(from the observation 
on day 55) for the 2-20 keV energy band. The upper
panel shows the Leahy normalized power with the Poisson noise level 
of 2 subtracted. For better display this has been logarithmically rebinned 
in frequency. At low frequencies the power spectrum is nearly flat, 
while at high frequencies it drops off as a power-law with index near -1. 
In the lower panel we plot the power minus Poisson noise level 
multiplied by the frequency, which is proportional to the 
source power per decade of frequency.  Evident in this plot
is a QPO near 0.5 Hz. 

We fit the power spectra in the 2-20 keV band with a model with 
three components representing the Poisson level, the continuum, 
and the QPO:
\begin{equation}
P = A+B[(f/f_0)^{\alpha}+(f/f_0)^{\beta}]^{-1}
   +C[1+4(f-f_{QPO})^2/(W_{QPO})^2]^{-1}~~.
\end{equation}
where $A$,$B$, and $C$ are the amplitudes of the Poisson level, 
continuum, and QPO components, $f_0$ the break point in the continuum 
power-law index, $\alpha$ and $\beta$ the continuum index before and after
the break, respectively, $f_{QPO}$ the QPO center, and $W_{QPO}$ the
QPO full-width at half maximum. In general this model reasonably 
characterized the power spectra. 

\begin{figure}[!t]
\centerline{\includegraphics[scale=.50]{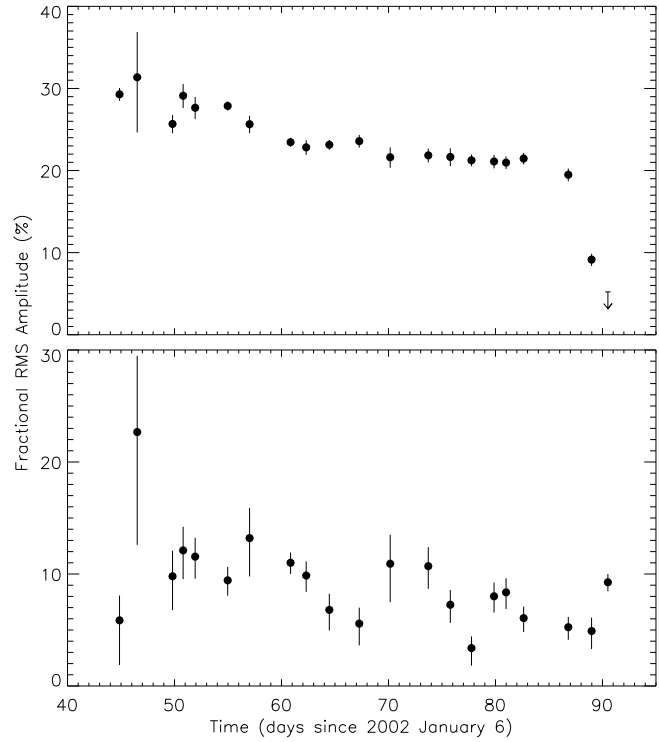}}
\vspace{0.2in}
\caption{\baselineskip =0.5\baselineskip
The time variation of the fractional rms amplitude of the 
continuum (upper panel) and QPO components (lower panel) 
in the 2$-$20 keV band. 
The indicated upper-limit is at the 2$\sigma$ level.\label{fig:xte_fitrms}}
\end{figure}

\begin{figure}[!t]
\centerline{\includegraphics[scale=.50]{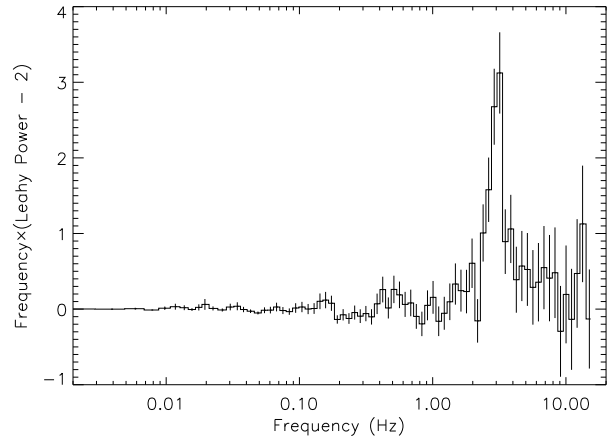}}
\vspace{0.2in}
\caption{\baselineskip =0.5\baselineskip
Power spectrum from an observation at the peak of the 2-5 keV flux
(day 90).\label{fig:xte_tpeak}}
\end{figure}

Figure \ref{fig:xte_fitrms} shows the resulting fractional rms amplitude 
of the continuum and QPO components during the first outburst. The continuum 
amplitude follows the trend described for Figure \ref{fig:xte_rmshist}. 
The QPO amplitude varies from 3\% to 13\%. As the 10-20 keV flux falls, 
the ratio of the QPO amplitude to the continuum 
amplitude increases. For the observation at the peak of the 2$-$5 keV flux, 
we have only an upper limit for the continuum RMS, which is below
the measured QPO amplitude. The power spectrum for this observation is shown in 
Figure \ref{fig:xte_tpeak}.

\begin{figure}[!t]
\centerline{\includegraphics[scale=.50]{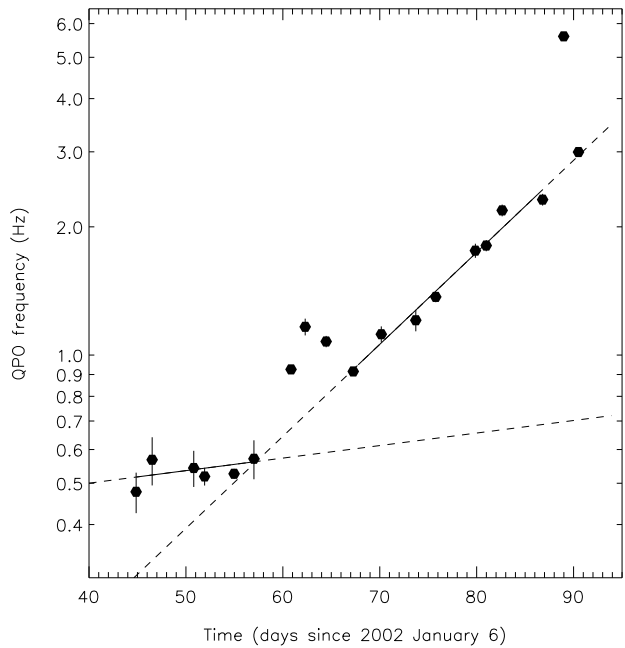}}
\vspace{0.2in}
\caption{\baselineskip =0.5\baselineskip
The variations in QPO frequency with time. The solid lines indicate
the intervals over which the logarithmic-linear fits were performed. 
The dashed lines are the extrapolations of each fit. \label{fig:xte_qpohist}}
\end{figure}

In Figure \ref{fig:xte_qpohist} we show the frequency evolution of the QPO. The 
frequency gradually increases during the early stages of the outburst and 
around day 60 it exhibits a jump, which coincides with the knee in the flux 
seen in Figure 2. After that the
QPO frequency rises more rapidly, following a different logarithmic trend.
The last two frequency measurements were during the state transition episode
of the outburst.

\section{Discussion} 


In the course of its X-ray activity XTE J1908+094 proceeds through a series
of X-ray states characteristic of black hole binaries, which we shall discuss 
using the terminology proposed by McClintock  \& Remillard (2003). The 
discovery outburst begins in the low/hard state, which lasted until about 
 day 87. The source then enters an intermediate state during which 
the 2$-$5 keV flux peaks. This ends near  day 90 
when the source enters the thermal-dominant (high/soft) state, which persists
until  day 143, just before the secondary peak, where the source enters an 
intermediate state and then after  day 149 returns to the low/hard
state. The remaining observations (in 1.5$-$12 keV) up to the peak of the 
second outburst show the source in the low/hard state (see Figure 
\ref{fig:rates_obss}). The one observation following 
the second outburst peak is consistent with the thermal-dominant state. 


In the low/hard state the energy spectra are dominated by a hard power-law 
component, and the power spectra by strong band-limited noise. During the 
low/hard state at the onset of the first outburst, the index of the power-law 
spectra began at $\Gamma=1.4$, and then gradually softened to $\Gamma=1.7$. The 
power spectra show band-limited noise with an rms amplitude which began near 
$r=30$\% and gradually fell to $r=22$\%. In addition there was a QPO with rms 
amplitude varying from 3\% to 13\%, which rose in frequency from 0.5 to 2.2 Hz. 
In the second interval of low/hard state ( days 149$-$295) the flux is lower,
and the behavior of the power-law index is more complex. Due to the low flux, 
we could not make significant power-spectral measurements.

Outburst onsets in the low/hard state have been seen in a number of X-ray novae.
Brocksopp et al. (2002) tabulate 13 sources with outbursts that began in the 
low/hard state, five of which never left this state. Strong low-frequency QPO 
with rising frequencies are common in these low/hard state onsets, and have 
been seen for GRO J0422+32 \cite{vanderHooft99}, GRO J1719-24 
\cite{vanderHooft96}, XTE J1550-564 (Finger et al. 1998; Cui et al. 1999), 
4U 1630-472 \cite{Dieters00}, XTE J1859+226 \cite{Markwardt99}, 
and XTE J1118+480 \cite{Wood00}, among others. 

\begin{figure}[!b]
\centerline{\includegraphics[scale=.50]{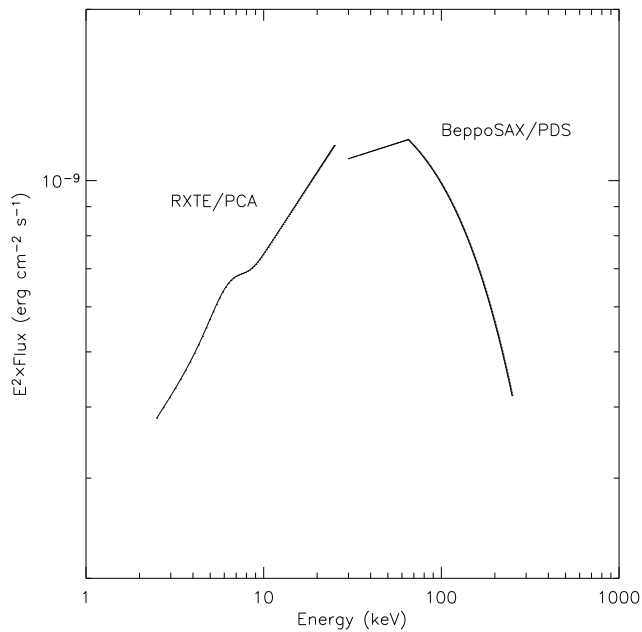}}
\vspace{0.0in}
\caption{\baselineskip =0.5\baselineskip
The RXTE/PCA spectral fit for day 64 shown together with the
BeppoSAX/PDA spectral fit for days 63$-$65 from in't Zand et al. (2002).
\label{fig:PDAspec}}
\end{figure}

Hard-X-ray and gamma-ray observations have shown that in the low/hard state the 
power-law spectra break in the 100 keV range \cite{Grove98}. In Figure 
\ref{fig:PDAspec} we show our spectral fit for the PCA data on day 64 along 
with the spectral fit for BeppoSAX/PDA data from days 62$-$65 
(MJD 52342$-$52345)
\cite{intZand02}. We notice a break near 50 keV. The flux in the 30-250 keV 
range is $3.2\times 10^{-9}~{\rm erg~cm}^{-2}~{\rm s}^{-1}$, which surpasses 
the flux in the 2.5-25 keV range of 
$2.63\times 10^{-9}~{\rm erg~cm}^{-2}~{\rm s}^{-1}$.

The low/hard state is also associated with radio emission. Flat spectrum radio 
emission, associated with compact jets, is consistently observed during the 
low/hard state of X-ray novae \cite{Fender03}. Indeed, during the onset of 
the first XTE J1908+094 outburst, Very Large Array observations 
(on days 74$-$75) 
led to the discovery of a radio counterpart to XTE J1908+094, with a flux of 
0.85 mJy at 8.6 GHz \cite{Rupen02a}. This was detected in additional 
observations until day 127 \cite{Rupen02b}. 


\begin{figure}[!b]
\centerline{\includegraphics[scale=.50]{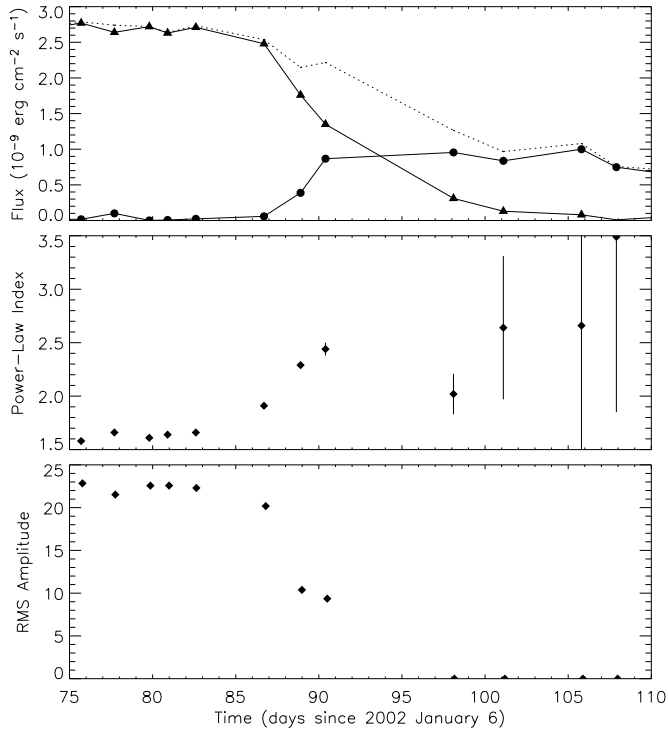}}
\vspace{0.0in}
\caption{\baselineskip =0.5\baselineskip
Flux, Power-law index, and variability evolution during the intermediate state.
The top panel shows the 2.5-25 keV flux for the power-law (filled triangles), the disk black-body component (filled circles), and their sum (dashed). 
The middle panel shows the power-law index. The lower 
panel shows the variability amplitude.\label{fig:IMstate}}
\end{figure}

Figure \ref{fig:IMstate} shows the X-ray flux, power-law index and variability 
amplitude evolution during the transition interval between the low/hard and 
thermal-dominant state. This intermediate state begins near day 67, 
when the power-law flux begins to drop, the power-law component begins to 
rapidly soften, and the disk black-body flux begins
to rise, and the flux variability begins to fall. The rise of the
disk black-body flux occurs in four days, but the fall of the power-law flux 
takes 15 days to complete. BeppoSAX/MECS observations covering day 66.4
$-$67.8 (MJD 52366.4$-$52367.8) show the onset of this transition \cite{intZand02}. 

In this transition, the total flux in 2.5$-$25 keV band consistently falls. 
However, a significant fraction of the disk black-body flux is below 
this energy range. From our 
spectral fits we find that the total integrated flux of this thermal component 
rises to $\sim 7\times 10^{-9}~{\rm erg~cm}^{-2}~{\rm s}^{-1}$,
implying that the bolometric flux may be constant or rising.


After day 90 the source is in the thermal-dominant state, with thermal disk 
flux dominating the spectrum, and low variability. The disk black-body 
normalizations average about 40, which is consistent with the inner disk 
being at the radius of the inner most stable circular orbit if
\begin{equation}
 (M/M_\odot)D_{10kpc}^{-1}\cos^{\onehalf}\theta \approx 4.5
\end{equation}
where $M$ is the black-hole mass,  $D_{10kpc}$ the source distance in units
of 10 kpc, and $\theta$ the disk inclination to the line of sight. Note here
that for any value of disk inclination angle, the mass of the central object 
is in the range of a black hole, if the source distance is of the order of 
10 kpc.

During this thermal-dominant state, the flux steadily falls. This mainly 
occurs by the temperature decreasing. The thermal-dominant state ends on  
day 149 when a transition begins back to the low/hard state.

 
In the transition from the low/hard state to the thermal-dominant state
starting day 87),
the disk black-body normalization begins near 15 and rises, implying an 
increasing inner disk radius. The opposite occurs on the transition back
to the low/hard state. This is counter to the expectation that the inner-disk 
radius is large during the low/hard state, and near the inner most stable orbit 
in the thermal-dominant state (e.g., Esin et al. 1997). This rise and fall may 
be due to systematic problems with our spectral fits: we detect 
only the high-energy tail of the thermal spectrum. In fits where the column 
density, disk temperature and flux are all free to vary, these parameters are, 
therefore, highly coupled. By fixing the column density to the value found with 
the BeppoSAX data, we have reduced this coupling, but could be biasing the 
solution.


There is a strong correlation between the flux associated with 
the broad line feature and that of the power law component during the 
early stages of the first outburst episode. One possible interpretation 
of this feature is that it is the fluorescent Fe K$\alpha$ emission produced
by the reprocessing of the hard X-ray photons by cooler material close to the
central object. The line centroid energies were somewhat lower 
than what is expected for neutral iron (6.4 keV). This may suggest that what we 
observe is primarily the red wing of Doppler shifted neutral Fe K$\alpha$ in 
a Keplerian accretion. This was seen also in 4U 1630$-$47 (Cui, Chen
\& Zhang 1999) and in XTE J1748$-$288 (Miller et al. 2001).


Esin et al. (1997) have presented a model to explain the states of X-ray novae. 
In the quiescent and low/hard state a thin accretion disk is present but 
truncated at a large inner radius. Within this radius there is an Advection 
Dominated Accretion Flow (ADAF) which is a hot and radiatively inefficient 
flow where most of the thermal energy generated is advected onto the black 
hole rather then being radiated. Above the accretion disk is a hot corona,  
which is a continuation of the advection dominated flow, which produces a
power-law spectral component in the low/hard state by Comptonization.

This model does not incorporate the jets which are responsible for the radio 
emission now known to be associated with the low/hard state. Markoff, Falcke 
\& Fender (2001) have proposed that these jets also produce the power-law 
component via synchrotron radiation. In their model a standard 
accretion disk transitions at an inner radius of $\sim 10^3$ km to a hot 
ADAF-like flow which feeds the jet. The hard X-rays are synchrotron radiation 
produced in a shock acceleration region some $10^3$ km above the disk plane. 
The radio emission is from beyond this region.

While providing successful fits of a multi-wavelength spectrum, neither of 
these models yet consider dynamical changes in the flow or attempt to explain 
the power-spectra seen in the different states. The high amplitude variability 
seen in the low-hard state requires changes in emissivity that are spatially 
coherent over most of the emission region. It is tempting to associate the 
QPO's seen in the low/hard state with the Keplerian  frequency at the inner 
edge of the thin accretion disk. Yet while this is plausible in the
low-hard state, the QPO in XTE J1908+094 persists into the intermediate state, 
where the inner disk radius inferred from spectral fits imply frequencies much 
larger than those observed.

\acknowledgments 

We thank the anonymous referee for helpful suggestions. E.G., M.H.F., C.K., P.M.W. 
acknowledge support from NASA grant NAG5-3370.

\clearpage

\begin{table}[!t] 
\caption{{\it RXTE} Observations of XTE J1908+094. \label{tbl:1}} 
 \small{
 \begin{center} 

\begin{tabular}{rcccc|rcccc} 
\hline 
\hline 
Obs  & Time\tablenotemark{a} & Exp  & Count Rate\tablenotemark{b} & Hardness  & Obs & Time\tablenotemark{a} & Exp  & Count Rate\tablenotemark{b} & Hardness   \\
\#   &                       & (ks) & (c/s/PCU)  & Ratio\tablenotemark{c}     & \#  &                   & (ks) & (c/s/PCU)  & Ratio\tablenotemark{c}  \\
\hline 
1   &   0.1 &  9.9 &  $<$0.09	    &  & 33  & 115.4 &  1.2 &  11.6  & 0.035(3) 	 \\	  
2   &   1.2 & 10.0 &  $<$0.09	    &  & 34  & 119.3 &  0.6 &  13.0  & 0.048(3)        \\	  
3   &   1.6 &  9.7 &  $<$0.04	    &  & 35  & 123.9 &  1.3 &  8.1   & 0.043(6) 	\\	  
4   &   2.1 & 10.2 &  $<$0.06	    &  & 36  & 125.2 &  0.8 &  7.6   & 0.034(8) 	 \\	  
5   &   5.2 &  9.8 &  $<$0.04	    &  & 37  & 129.9 &  0.9 &  8.7   & 0.058(5) 	 \\	  
6   &  11.1 & 10.0 &  $<$0.02	    &  & 38  & 131.1 &  1.6 &  6.4   & 0.037(3) 	\\	  
7   &  44.9 &  9.4 &  12.1  & 0.224(3) & 39  & 137.4 &  1.1 &  3.8   & 0.042(8)    \\ 
8   &  46.4 &  3.0 &  13.4  & 0.199(3) & 40  & 141.3 &  1.1 &  6.2   & 0.080(6)     \\        
9   &  49.7 &  2.9 &  15.7  & 0.223(2) & 41  & 144.2 &  1.6 &  7.0   & 0.090(4)     \\        
10  &  50.7 &  2.9 &  16.5  & 0.226(2) & 42  & 145.5 &  1.7 &  6.8   & 0.099(4)     \\        
11  &  51.9 &  2.9 &  18.1  & 0.201(3) & 43  & 149.5 &  0.9 &  7.4   & 0.132(4)     \\        
12  &  54.9 &  3.3 &  21.4  & 0.207(1) & 44  & 154.6 &  1.1 &  5.4   & 0.182(8)     \\        
13  &  56.9 &  3.3 &  24.3  & 0.207(1) & 45  & 160.9 &  0.7 &  5.0   & 0.169(8)    \\ 
14  &  60.7 &  8.9 &  29.6  & 0.215(1) & 46  & 163.2 &  0.6 &  4.8   & 0.177(9)     \\        
15  &  62.2 &  2.1 &  31.5  & 0.212(1) & 47  & 170.5 &  1.3 &  3.5   & 0.165(9)     \\        
16  &  64.4 &  1.5 &  32.3  & 0.209(1) & 48  & 174.2 &  1.1 &  3.7   & 0.171(8)    \\ 
17  &  67.2 &  1.2 &  33.3  & 0.215(2) & 49  & 179.4 &  1.0 &  3.4   & 0.122(8)    \\	
18  &  70.1 &  3.1 &  33.8  & 0.208(1) & 50  & 183.3 &  0.9 &  3.1   & 0.114(7)    \\	
19  &  73.6 &  2.4 &  35.3  & 0.194(1) & 51  & 188.9 &  1.6 &  3.2   & 0.147(8)    \\	
20  &  75.8 &  0.9 &  36.6  & 0.190(2) & 52  & 192.4 &  0.7 &  2.9   & 0.090(18)    \\         
21  &  77.8 &  1.4 &  35.3  & 0.193(1) & 53  & 194.2 &  1.2 &  2.6   & 0.107(15)    \\         
22  &  79.9 &  3.4 &  35.8  & 0.200(1) & 54  & 198.8 &  1.8 &  2.6   & 0.181(9)    \\	
23  &  80.9 &  3.2 &  35.1  & 0.193(1) & 55  & 207.6 &  1.1 &  2.2   & 0.133(10)     \\   
24  &  82.7 &  1.9 &  37.8  & 0.175(1) & 56  & 238.9 &  0.9 &  2.5   & 0.158(20)     \\   
25  &  86.8 &  1.9 &  39.0  & 0.139(1) & 57  & 244.0 &  1.2 &  2.4   & 0.148(11)    \\         
26  &  88.9 &  2.5 &  38.1  & 0.084(1) & 58  & 251.8 &  0.8 &  2.6   & 0.159(19)     \\   
27  &  90.5 &  1.6 &  41.6  & 0.048(1) & 59  & 257.5 &  1.9 &  2.6   & 0.151(13)     \\   
28  &  98.2 &  1.4 &  26.2  & 0.029(1) & 60  & 269.0 &  1.9 &  2.2   & 0.177(14)    \\         
29  & 101.2 &  1.5 &  20.7  & 0.018(1) & 61  & 283.1 &  1.6 &  2.7   & 0.163(9)   \\  
30  & 105.9 &  1.9 &  15.3  & 0.013(1) & 62  & 295.0 &  1.2 &  2.8   & 0.172(9)   \\  
31  & 107.9 &  0.9 &  16.0  & 0.005(2) & 63  & 339.1 &  1.4 &  11.4  & 0.203(3)   \\	
32  & 112.9 &  3.1 &  13.7  & 0.019(1) & 64  & 384.7 &  1.5 &  10.1  & 0.032(2)   \\	       
\hline				  

\end{tabular} 
\end{center} 
}
\begin{flushleft}
a$-$ Observation times referenced to 2002 January 6.0 (MJD 52280.0).

b$-$ Background subtracted, orbit averaged rates in 2$-$20 keV band.

c$-$ Hardness Ratios defined as the ratio of the background subtracted, 
orbit averaged rates in the 10$-$20 keV to those in 2$-$10 keV.
\end{flushleft}

\end{table}

\clearpage

\begin{deluxetable}{cccc|ccc|cc|c}
\tablewidth{0pc}
\tabletypesize{\footnotesize}
\tablecaption{Spectral fit parameters. \label{tbl:2}} 
\tablehead{ 
\colhead{Obs} & \multicolumn{3}{c|}{Multicolor Disk Blackbody} & 
\multicolumn{3}{c|}{Broad Line} & \multicolumn{2}{c|}{Power Law} & \multicolumn{1}{c}{ } \\ 
\colhead{\#}  & \multicolumn{1}{c}{kT$_{\rm disk}$\tablenotemark{a}} & \multicolumn{1}{c}{N$_{\rm disk}$} & 
\multicolumn{1}{c|}{F$_{\rm disk}$\tablenotemark{b}} & \multicolumn{1}{c}{E$_{\rm cent.}$\tablenotemark{a}} & 
\multicolumn{1}{c}{$\sigma$\tablenotemark{a}} & \multicolumn{1}{c|}{F$_{\rm line}$\tablenotemark{b}} & 
\multicolumn{1}{c}{$\Gamma$} & \multicolumn{1}{c|}{F$_{\rm PL}$\tablenotemark{b}} & F$_{\rm Total}$\tablenotemark{b}  \\
\colhead{ }   & \multicolumn{1}{c}{(keV)} & \multicolumn{1}{c}{} & 
\multicolumn{1}{c|}{(10$^{-10}$cgs)} & \multicolumn{1}{c}{(keV)} & 
\multicolumn{1}{c}{(keV)} & \multicolumn{1}{c|}{(10$^{-11}$cgs)} & \multicolumn{1}{c}{ } &
\multicolumn{1}{c|}{(10$^{-9}$cgs)} & (10$^{-9}$cgs)
}
\startdata
7   &  0.60	     &  $<$11.5 	&  $<$0.03   &  5.72$\pm$0.17 & 1.07$\pm$0.55 & 2.21(44) & 1.40$\pm$0.03 & 1.09(1) & 1.12(2)  \\
8   &  0.60	     &  $<$39.5 	&  $<$0.09   &  5.75$\pm$0.27 & 1.03$\pm$0.92 & 1.92(66) & 1.45$\pm$0.04 & 1.15(2) & 1.18(4)  \\
9   &  0.60	     &  $<$149.7	&  $<$0.35   &  5.44$\pm$0.14 & 0.96$\pm$0.44 & 2.76(41) & 1.50$\pm$0.03 & 1.30(1) & 1.37(2)  \\
10  &  0.60	     &  $<$170.0	&  $<$0.40   &  5.49$\pm$0.13 & 0.96$\pm$0.43 & 2.88(41) & 1.49$\pm$0.02 & 1.36(1) & 1.44(2)  \\
11  &  0.60	     &  $<$0.01 	&  $<$0.01   &  5.44$\pm$0.43 & 1.10$\pm$1.35 & 1.44(75) & 1.45$\pm$0.03 & 1.57(2) & 1.59(5)  \\
12  &  0.60	     &  $<$47.1 	&  $<$0.11   &  5.33$\pm$0.11 & 1.16$\pm$0.34 & 2.81(32) & 1.47$\pm$0.01 & 1.83(1) & 1.87(2)  \\
13  &  0.60	     &  $<$31.6 	&  $<$0.07   &  5.54$\pm$0.12 & 1.25$\pm$0.43 & 2.46(37) & 1.50$\pm$0.01 & 2.06(1) & 2.09(2)  \\
14  &  0.60	     &  $<$76.9 	&  $<$0.02   &  5.60$\pm$0.15 & 1.14$\pm$0.50 & 3.36(55) & 1.55$\pm$0.02 & 2.34(2) & 2.39(3)  \\
15  &  0.60	     &  $<$0.01 	&  $<$0.01   &  5.68$\pm$0.11 & 1.31$\pm$0.37 & 5.34(55) & 1.52$\pm$0.01 & 2.57(2) & 2.63(4)  \\
16  &  0.60	     &  $<$37.2 	&  $<$0.08   &  5.61$\pm$0.21 & 1.14$\pm$0.69 & 2.68(65) & 1.55$\pm$0.02 & 2.61(2) & 2.65(3)  \\
17  &  0.60	     &  $<$0.01 	&  $<$0.01   &  5.59$\pm$0.23 & 1.10$\pm$0.76 & 3.32(82) & 1.53$\pm$0.02 & 2.69(3) & 2.73(6)  \\
18  &  0.60	     &  $<$78.3 	&  $<$0.18   &  5.71$\pm$0.08 & 1.11$\pm$0.29 & 4.34(38) & 1.54$\pm$0.01 & 2.66(1) & 2.73(2)  \\
19  &  0.60	     &  $<$68.1 	&  $<$0.16   &  5.41$\pm$0.09 & 1.15$\pm$0.28 & 4.44(37) & 1.57$\pm$0.01 & 2.69(1) & 2.76(2)  \\
20  &  0.60	     &  $<$69.3 	&  $<$0.17   &  5.77$\pm$0.18 & 1.12$\pm$0.62 & 4.02(78) & 1.58$\pm$0.02 & 2.77(3) & 2.83(5)  \\
21  &  0.60	     &  $<$0.01 	&  $<$0.01   &  5.78$\pm$0.14 & 1.12$\pm$0.49 & 3.99(62) & 1.66$\pm$0.01 & 2.64(2) & 2.68(4)  \\
22  &  0.60	     &  $<$0.01 	&  $<$0.01   &  5.89$\pm$0.09 & 1.27$\pm$0.32 & 4.84(46) & 1.61$\pm$0.01 & 2.72(1) & 2.78(2)  \\
23  &  0.60	     &  $<$29.6 	&  $<$0.07   &  5.80$\pm$0.08 & 1.11$\pm$0.30 & 4.28(40) & 1.64$\pm$0.01 & 2.63(1) & 2.68(2)  \\
24  &  0.60	     &  $<$97.8 	&  $<$0.23   &  5.83$\pm$0.10 & 1.10$\pm$0.35 & 4.80(50) & 1.66$\pm$0.01 & 2.71(2) & 2.79(3)  \\
25  &  0.75	     &  $<$55.4 	&  $<$0.58   &  6.08$\pm$0.11 & 1.10$\pm$0.39 & 4.73(56) & 1.91$\pm$0.02 & 2.48(2) & 2.59(4)  \\
26  &  0.79$\pm$0.01  &  276$\pm$27      &  3.88(37) &  5.95$\pm$0.07 & 1.10$\pm$0.33 & 4.91(52) & 2.29$\pm$0.03 & 1.76(2) & 2.19(7)  \\
27  &  0.74$\pm$0.01  &  914$\pm$55      &  8.67(52) &  5.96$\pm$0.37 & 1.20$\pm$0.48 & 4.40(70) & 2.44$\pm$0.06 & 1.35(2) & 2.27(9)  \\
28  &  0.69$\pm$0.01  &  1455$\pm$69  	 &  9.55(45)  &  6.0	      & 1.15$\pm$0.55 & $<$2.80    & 2.03$\pm$0.19 & 0.31(3) & 1.29(8)  \\
29  &  0.67$\pm$0.01  &  1696$\pm$84   	 &  8.37(42)  &  6.0	      & 0.67$\pm$1.55 & $<$0.83    & 2.64$\pm$0.67 & 0.13(2) & 0.98(7)  \\
30  &  0.69$\pm$0.01  &  1686$\pm$99  	 &  10.0(59)  &  6.0	      & 1.10	      & $<$0.41    & 2.66$\pm$1.64 & 0.08(2) & 1.09(11)  \\
31  &  0.63$\pm$0.01  &  2139$\pm$193    &  7.48(68)  &  6.0	      & 0.62$\pm$1.14 & $<$0.81    & 3.49$\pm$1.64 & 0.01(3) & 0.76(7) \\
32  &  0.65$\pm$0.01  &  1456$\pm$76  	 &  5.96(31)  &  6.0	      & 1.07$\pm$2.10 & $<$0.77    & 3.03$\pm$1.22 & 0.08(0) & 0.68(7)  \\
33  &  0.61$\pm$0.01  &  2045$\pm$188  	 &  5.30(49)  &  6.0	      & 1.00$\pm$0.62 & $<$2.05    & 0.76$\pm$0.51 & 0.10(2) & 0.66(7)  \\
34  &  0.59$\pm$0.01  &  1920$\pm$251    &  4.24(55)  &  6.0	      & 0.67$\pm$2.89 & $<$0.76    & 2.35$\pm$0.38 & 0.29(3) & 0.73(10) \\
35  &  0.61$\pm$0.01  &  1308$\pm$253  	 &  3.60(70)  &  6.0	      & 1.14$\pm$0.71 & $<$2.34    & 3.37$\pm$0.76 & 0.08(4) & 0.47(12) \\
36  &  0.54$\pm$0.01  &  2713$\pm$803    &  3.21(95)  &  6.0	      & 1.10 	      & $<$0.20    & 2.58$\pm$1.84 & 0.10(3) & 0.43(14) \\
37  &  0.58$\pm$0.02  &  1189$\pm$303    &  2.24(57)  &  6.0	      & 1.15$\pm$1.97 & $<$1.36    & 2.54$\pm$0.51 & 0.25(3) & 0.49(11) \\
38  &  0.54$\pm$0.02  &  2024$\pm$461	 &  2.39(54)  &  6.0	      & 1.36$\pm$2.57 & $<$0.84    & 2.90$\pm$0.97 & 0.07(3) & 0.32(10) \\
39  &  0.57$\pm$0.02  &  1042$\pm$263  	 &  1.73(44)  &  6.0	      & 0.88$\pm$2.16 & $<$1.01    & 2.54$\pm$1.44 & 0.08(3) & 0.19(9)  \\
40  &  0.57$\pm$0.04  &  686$\pm$297     &  1.14(50)  &  6.0	      & 1.43$\pm$2.39 & $<$1.18    & 2.36$\pm$0.39 & 0.23(3) & 0.36(11) \\
41  &  0.68$\pm$0.03  &  178$\pm$59      &  0.99(33)  &  6.0	      & 1.10$\pm$1.14 & $<$1.59    & 2.16$\pm$0.26 & 0.31(2) & 0.43(7)  \\
42  &  0.64$\pm$0.05  &  163$\pm$89      &  0.61(33)  &  6.0	      & 1.32$\pm$2.78 & $<$0.76    & 2.26$\pm$0.23 & 0.35(2) & 0.42(8)  \\
43  &  0.60	     &  $<$69.9 	&  $<$0.17   &  5.97$\pm$0.35 & 1.09$\pm$1.39 & 1.15(66) & 1.89$\pm$0.08 & 0.48(2) & 0.50(3)  \\
44  &  0.60	     &  $<$37.8 	&  $<$0.10   &  5.43$\pm$0.29 & 1.10$\pm$0.93 & 1.78(62) & 1.51$\pm$0.14 & 0.39(2) & 0.42(3)  \\
45  &  0.60	     &  $<$35.7 	&  $<$0.09   &  5.59$\pm$0.32 & 1.08$\pm$1.03 & 1.52(60) & 1.46$\pm$0.15 & 0.34(2) & 0.37(3)  \\
46  &  0.60	     &  $<$40.3 	&  $<$0.09   &  6.52$\pm$0.65 & 1.10	     & 0.35(46) & 1.45$\pm$0.13 & 0.35(1) & 0.36(3)  \\
47  &  0.60	     &  $<$0.01 	&  $<$0.001  &  6.0	      & 1.10	    & $<$0.01	 & 1.68$\pm$0.06 & 0.26(1) & 0.26(1)  \\
48  &  0.60	     &  $<$19.0 	&  $<$0.04   &  6.0	      & 1.10	    & $<$0.44	 & 1.53$\pm$0.12 & 0.27(1) & 0.28(2)  \\
49  &  0.60	     &  $<$0.01 	&  $<$0.001  &  6.0	      & 1.10	    & $<$0.32	 & 1.86$\pm$0.16 & 0.22(1) & 0.22(2)  \\
50  &  0.60	     &  $<$61 1 	&  $<$0.14   &  6.0	      & 1.10	    & $<$0.25	 & 2.04$\pm$0.17 & 0.19(2) & 0.20(4)  \\
51  &  0.60	     &  $<$17.3 	&  $<$0.04   &  6.12$\pm$1.22 & 1.06$\pm$1.58 & 2.12(87) & 1.94$\pm$0.18 & 0.24(2) & 0.25(4)  \\
52  &  0.60	     &  $<$0.01 	&  $<$0.01   &  6.46$\pm$0.58 & 1.41$\pm$2.08 & 2.31(1.84) & 2.20$\pm$0.46 & 0.19(5) & 0.21(12) \\
53  &  0.60	     &  $<$61.1 	&  $<$0.15   &  6.67$\pm$0.40 & 1.09$\pm$1.83 & 1.69(1.22) & 1.88$\pm$0.35 & 0.18(3) & 0.20(7)  \\
54  &  0.60	     &  $<$23.5 	&  $<$0.06   &  5.18$\pm$0.56 & 0.82$\pm$1.59 & 0.64(44) & 1.33$\pm$0.20 & 0.22(1) & 0.24(2)  \\
55  &  0.60	     &  $<$9.84 	&  $<$0.02   &  6.02$\pm$0.45 & 1.82$\pm$1.66 & 1.53(83) & 1.89$\pm$0.21 & 0.15(1) & 0.16(3)  \\
56  &  0.60	     &  $<$37.7 	&  $<$0.09   &  6.12$\pm$1.74 & 1.10	      & $<$0.01    & 1.46$\pm$0.35 & 0.24(2) & 0.25(2)  \\
57  &  0.60	     &  $<$0.01 	&  $<$0.01   &  6.09$\pm$0.80 & 1.24$\pm$3.74 & 0.61(99) & 1.74$\pm$0.23 & 0.21(2) & 0.22(5)  \\
58  &  0.60	     &  $<$25.2 	&  $<$0.06   &  5.79$\pm$1.66 & 1.10	      & 0.34(67) & 1.53$\pm$0.40 & 0.22(2) & 0.23(4)  \\
59  &  0.60	     &  $<$0.01 	&  $<$0.01   &  7.01$\pm$1.74 & 1.10	      & 0.27(80) & 1.76$\pm$0.22 & 0.19(2) & 0.20(5)  \\
60  &  0.60	     &  $<$45.7 	&  $<$0.11   &  5.32$\pm$0.53 & 1.1$\pm$1.41  & 1.04(57) & 1.94$\pm$0.61 & 0.19(2) & 0.21(4)  \\
61  &  0.60	     &  $<$3.46 	&  $<$0.08   &  5.82$\pm$0.40 & 1.08$\pm$1.62 & 1.05(69) & 1.56$\pm$0.20 & 0.24(2) & 0.25(4)  \\
62  &  0.60	     &  $<$0.01 	&  $<$0.01   &  6.88$\pm$0.86 & 1.10	      & 0.38(45) & 1.55$\pm$0.12 & 0.31(1) & 0.32(3)  \\
63  &  0.60	     &  $<$15.0 	&  $<$0.04   &  5.63$\pm$0.43 & 1.28$\pm$1.72 & 1.13(77) & 1.43$\pm$0.04 & 1.19(2) & 1.20(4)  \\
64  &  0.60$\pm$0.01   &  1669$\pm$144    &  4.40(38)  &  6.23$\pm$0.33 & 1.08$\pm$0.87 & 1.43(48) & 1.34$\pm$0.64 & 0.08(2) & 0.54(7)  \\
\enddata
\begin{flushleft}
a$-$ Values without errors were kept frozen in the fitting procedure.

b$-$ Absorbed flux in the 2.5$-$25 keV band. Values in the parentheses are the 
2-sigma level errors in the last number of digits of given flux values.
\end{flushleft}

\end{deluxetable}


\begin{thebibliography}{}

\bibitem[Brocksopp et al. 2002]{Brocksopp02}
   Brocksopp, C., Fender, R. P., McCollough, M. et al.  2002, MNRAS, 331, 765
\bibitem[Brocksopp et al.  2001]{Brocksopp01}
   Brocksopp, C., Jonker, P. G., Fender, R. P. et al. 2001, MNRAS, 323, 517
\bibitem{cannizzo}
   Cannizzo, J.K. (1993), \apj, 419, 318
\bibitem{chaty/1908}
   Chaty, S., Mignani, R.P. and Israel, G.L. 2002, MNRAS 337, L23
\bibitem{chen}
   Chen, W., Shrader, C.R. and Livio, M. 1997, ApJ 491, 312
\bibitem[Cui et al.  1999]{Cui99}
   Cui, W., Zhang, S. N., Chen, W. \& Morgan E. H. 1999, \apj, 512, L43
\bibitem[Cui et al.  2000]{Cui00}
   Cui, W., Chen, W. \& Zhang, S. N. 2000, \apj, 529, 952
\bibitem[Dieters et al.  2000]{Dieters00}
   Dieters, S. W., Belloni, T., Kuulers, E. et al.  2000, \apj 538, 307
\bibitem{s1.5/dubus}
   Dubus, G., Hameury, J.-M and Lasota, J.-P. 2001, A\&A 373, 251
\bibitem[Esin, McClintock, \& Narayan 1997]{Esin97}
   Esin, A. A., McClintock, J. E., \& Narayan, R. 1997, \apj, 489, 865
\bibitem[Fender 2003]{Fender03}
   Fender, R. 2003, astro-ph/0303339
\bibitem[Finger et al. 1998]{Finger98}
   Finger, M. H., Dieters, S. W., \& Wilson, R. B. 1988, IAUC 7010
\bibitem{garna02}
   Garnavich, P., Quinn, J., Callanan, P. 2002, IAUC, 7877
\bibitem[Grove et al.  1998]{Grove98} 
   Grove, J. E., Johnson, W. N., Kroeger, R. A., et al. 1998, \apj 500, 899
\bibitem{homan/j1550}
   Homan, J., Wijnands, R., van der Klis, M., et al. 2001, ApJS, 132, 377
\bibitem{hulleman}
   Hulleman, F., Tennant, A. F., van Kerkwijk, M. H. et al. 2001, \apj, 563, L49
\bibitem{miller01}
   Miller, J.M., Fox, D.W., Di Matteo, T., et al. 2001, \apj, 546, 1055
\bibitem[van der Hooft et al. 1996]{vanderHooft96}
   van der Hooft, F., Kouveliotou, C., van Paradijs, J. et al.  1996,
   A\&A Supp., 120C, 141
\bibitem[van der Hooft et al. 1999]{vanderHooft99}
   van der Hooft, F., Kouveliotou, C., van Paradijs, J. et al.  1999,
   \apj, 513, 477
\bibitem[McClintock \& Remillard 2003]{McClintock03} 
   McClintock, J. E. \& Remillard, R. A. 2003, in Compact Stellar X--ray
   sources, eds. W.H.G. Lewin and M. van der Klis, in press, astro-ph/0306213
\bibitem[Markoff, Falcke, \& Fender 2001]{Markoff01}
   Markoff, S., Falcke, H, \& Fender, R. 2001, A\&A 372, L25
\bibitem[Markwardt et al. 1999]{Markwardt99}
   Markwardt, C. B., Focke, W. B., Swank, J. H. \& Taam, R. E. 1999,
   Bull. Am. Astron. Soc., 31, 1555
\bibitem{mitsuda}
   Mitsuda, K., Inoue, H., Koyama, K., et al. 1984, PASJ 36, 741
\bibitem[Rupen et al. 2002a]{Rupen02a}
   Rupen, M. P., Dhawan, V., \& Mioduszewski, A. J. 2002a, IAUC 7874
\bibitem[Rupen et al. 2002b]{Rupen02b}
   Rupen, M. P., Dhawan, V., \& Mioduszewski, A. J. 2002b, IAUC 8029
\bibitem{tanakalewin}
   Tanaka, Y. and Lewin, W.H.G. 1995, in X-ray Binaries, eds. W.H.G. Lewin, 
   J. van Paradijs, and E.P.J. van den Heuvel, (Cambridge U. Press, Cambridge),
   126
\bibitem{valinia}
   Valinia, A. \& Marshall, F.E. 1998, \apj, 505, 134
\bibitem[Wood et al.  2000b]{Wood00}
   Wood, K. S., Ray, P. S., Bandyopadhyay, R. M. et al.  2000,
   \apj, 544, L45
\bibitem{woods02}
   Woods, P. M., Kouveliotou, C., Finger, M. et al. 2002, IAUC 7856   
\bibitem[in't Zand et al. 2002]{intZand02} 
   in't Zand, J. J. M., Miller, J. M., Oosterbroek, T. \& Parmar, A. N. 2002, 
   A\&A 394, 553
 
\end{thebibliography}
\end{document}